\documentclass{aa}  

\usepackage{graphicx}
\usepackage[colorlinks=true, allcolors=blue]{hyperref}
\hypersetup{
     colorlinks   = true,
     citecolor    = blue,
     urlcolor = blue
}
\usepackage{txfonts}
\usepackage{lipsum}
\usepackage{tabularx}
\usepackage{subcaption}         
\usepackage{lscape}             
\usepackage{placeins}           
                                
\begin{document}

   \title{Fast radio burst -- persistent radio source systems}

   \subtitle{III. The relation between PRS luminosity and FRB rotation measure}

   \author{
          D. Pelliciari
          \inst{1}
          \and
          G. Bernardi\inst{1,2,3}
          \and
          L. Bruno\inst{1}
          \and
          M. Pilia\inst{4}
          \and
          P. Esposito\inst{5,6}
          \and
          L. Beduzzi\inst{1,7}
          \and
          A. Geminardi\inst{4,6,8}
          \and
          O. Smirnov\inst{2,3,1,9,10}
          }

   \institute{INAF-Istituto di Radio Astronomia (IRA), via Piero Gobetti 101, Bologna, Italy\\
              \email{davide.pelliciari@inaf.it}
         \and
             Centre for Radio Astronomy Techniques and Technologies (RATT), Department of Physics and Electronics, Rhodes University, Makhanda 6140, South Africa
         \and
             South African Radio Astronomy Observatory, Black River Park, 2 Fir Street, Observatory, Cape Town, 7925, South Africa
        \and
            INAF-Osservatorio Astronomico di Cagliari, via della Scienza 5, I-09047, Selargius (CA), Italy
        \and
            Scuola Universitaria Superiore IUSS Pavia, Palazzo del Broletto, piazza della Vittoria 15, I-27100 Pavia, Italy
        \and
            INAF–Istituto di Astrofisica Spaziale e Fisica Cosmica di Milano, via Corti 12, I-20133 Milano, Italy
        \and
            Dipartimento di Fisica e Astronomia, Universitá di Bologna, via Gobetti 93/2, 40129 Bologna, Italy
        \and
            Dipartimento di Fisica, Università di Trento, via Sommarive 14, I-38123 Povo (TN), Italy
        \and
            Astrophysics, Department of Physics, University of Oxford, Keble Road, Oxford, OX1 3RH, UK
        \and
            Breakthrough Listen, Astrophysics, Department of Physics, The University of Oxford, Keble Road, Oxford, OX1 3RH, UK
        }
   \date{Received April XX, 2026}
   
 
  \abstract
{Fast radio bursts (FRBs) are millisecond-duration radio transients of extragalactic origin whose physical origin remains uncertain. A small fraction of FRBs are known to repeat, and some of them are associated with persistent radio sources (PRSs), interpreted as synchrotron-emitting nebulae surrounding the FRB source. In the context of magnetar-based models, the rotation measure (RM) of an FRB is expected to correlate with the spectral luminosity of its associated PRS, providing a probe of the physical properties and evolution of the nebula.}
{We investigate the relation between FRB RM and PRS spectral luminosity, constrain the characteristic size of the PRS nebulae, and use the intrinsic scatter of the relation to investigate evolutionary scenarios for FRB--PRS systems.}
{We analyse a sample of 50 FRB sources with known RM, including both PRS detections and spectral luminosity upper limits, using a Bayesian MCMC framework that jointly accounts for detections and non-detections. We model the luminosity--RM relation as $L_\nu \propto \zeta_e \gamma_c^2 R^2 |{\rm RM}|^\beta$, constraining the characteristic nebular size $R$ and the RM scaling index $\beta$.}
{For the full sample, we obtain $R = 0.016^{+0.115}_{-0.014}$ pc at $1\sigma$ confidence for a free $\beta$, while fixing the RM dependence to the canonical linear scaling, $\beta=1$, gives $R = 0.008^{+0.058}_{-0.007}$ pc. The data mildly favour super linear RM scalings, although the inferred values of $\beta$ remain consistent with $\beta=1$ within $2\sigma$. We find a substantial intrinsic scatter in the luminosity--RM relation, significantly larger than previous estimates based on confirmed PRSs alone.}
{Interpreting this scatter within evolutionary models, our results favour scenarios involving efficient particle acceleration and/or rapid nebular expansion. Forward-shock models in supernova-remnant/interstellar-medium or pulsar-wind-nebula/supernova-remnant environments are broadly consistent with the observations, whereas models predicting slower evolution, such as pulsar-wind-nebula bubble and Sedov--Taylor scenarios, are disfavoured.}

   \keywords{methods: data analysis -- methods: statistical -- methods: observational -- techniques: interferometric -- stars: magnetars -- radio continuum: general}

   \maketitle
    \nolinenumbers
\section{Introduction}\label{sec:Intro}
Fast radio bursts \citep[FRBs; e.g.,][]{cordesChatterjee19,petroff21,Zhang22_rev} are cosmological radio transients of millisecond duration characterized by high radio luminosities and brightness temperatures, suggesting coherent emission from compact objects. Among the leading progenitor candidates, magnetars, i.e. neutron stars powered by strong internal magnetic fields \citep[see, e.g.,][for a recent review on the topic]{Rea25} are favored, and strongly supported by the 2020 detection of an FRB-like event from the Galactic magnetar SGR J1935+2154 \citep{CHIME20b,Bochenek20a,Mereghetti_July2021}, as well as from previous theoretical studies \citep[e.g.][]{Popov13,Liubarsky20,Beloborodov19,Sobacchi22}. A small fraction ($\sim 2\%$) of the overall FRB population has been observed to repeat \citep{CHIMECat2}, but nowadays it is still not clear whether the latter represent a different type of sources with respect to one-offs \citep[e.g.,][]{pleunis21,James23,sand24}. 

No prompt multiwavelength counterpart has been found for FRBs, despite substantial observational efforts toward their identification \citep[see][for a review]{Zhang24_multi}. Nevertheless, persistent radio sources (PRSs) co-located with a handful of precisely localized FRBs have been reported \citep{Marcote17,Niu21,Bruni23,Bruni24,Ibik24,Moroianu26,Mfulwane26}. PRSs are compact \citep[sub-parsec to parsec scales, e.g.,][]{Marcote17} and luminous \citep[$L_\nu \simeq 10^{29}$ erg s$^{-1}$ Hz$^{-1}$;][]{Law22}, continuum radio sources, whose emission cannot be explained in terms of star-formation activity at the FRB site. To date, four repeating FRBs have been  associated with a PRS: FRB 20121102A \citep[R1;][]{Chatterjee17,Marcote17}, FRB 20190520B \citep[R1-twin;][]{Niu21,Bhandari23b}, FRB 20190417A \citep{Ibik24,Moroianu26}, and FRB 20240114A \citep{Bruni24,Bhusare25}. These persistent sources share common properties such as, e.g., their localization in dwarf and star-forming galaxies \citep[see, e.g.,][]{Moroianu26}, flat or mildly inverted spectral indices \citep{Moroianu26,Bruno26} and large values for the rotation measure (RM) of the associated FRBs \citep[e.g.][]{Michilli18,AnnaThomas23,Moroianu26}.

Furthermore, a number of PRS candidates have been reported, associated with both repeating FRBs, namely FRBs 20201124A \citep{Bruni23} and 20181030A \citep{Ibik24}, and apparently one-off sources \citep[see ][]{Mfulwane26}. However, their physical sizes remain unconstrained at parsec scales and, in several cases, the FRB localization regions are too large to firmly establish a physical association with the candidate PRSs. In a companion work \citep[hereafter Paper I;][]{Pelliciari26_PaperI}, we have recently conducted a systematic search for PRS candidates using new $1.26$ GHz uGMRT observations of 24 FRBs, combined with literature data. In another companion work \citep[hereafter Paper II;][]{Pelliciari26_PaperII}, we followed up the candidate PRS associated with FRB 20181030A with EVN and eMERLIN observations, providing evidence for its compactness on sub-parsec scales.

Nowadays, the physical origin of PRSs remains debated. The most natural interpretation is synchrotron emission from a magnetar wind nebula (MWN): a compact magnetized nebula inflated by outflows from a young, active magnetar, powered either by the decay of its internal magnetic field or by spin-down \citep{margalitmetzger18,Rahaman26}.

Interestingly, a positive, linear correlation between the specific luminosity ($L_\nu$) of the PRS and the RM of the associated FRB bursts is expected if the persistent radio emission and the Faraday rotation originate from the same magnetized nebula surrounding the FRB engine \citep{Yang20,Yang22}. This $L_\nu - {\rm RM}$ relation has been derived by \citet{Yang20,Yang22}, who showed that it follows from this single assumption, regardless of the nature of the central engine. Hereafter, we refer to this relation as the Yang-Li-Zhang (YLZ) relation, consistently with, e.g., \cite{Gao25}. This scaling is naturally expected in the MWN framework described in \citet{Rahaman26}: in a one-zone MWN framework, RM scales linearly with the source spectral luminosity, in agreement with \citet{Yang20}. Whether the hypernebula model \citep{Sridhar22,Sridhar24} makes an analogous prediction is less straightforward. While in that framework both $L_\nu$ and ${\rm RM}$ are produced by the same nebular plasma, the two quantities depend on a large number of coupled free parameters and evolve on different timescales, such that the sign and slope of any resulting $L_\nu - {\rm RM}$ relation depend critically on which parameters are held fixed and which are allowed to vary. The hypernebula model is therefore best regarded as consistent with -- rather than predictive of -- the observed relation.

From an observational point of view, FRB-PRS systems agree well, although with some level of scatter \citep[see][]{Yang26}, with the prediction of compact nebulae \citep[see, e.g.,][]{Ibik24}. On the other hand, FRBs showing low or negligible Faraday rotation are expected to have correspondingly faint or absent PRSs. This is consistent with the non-detection of PRSs in several well-monitored, active repeaters such as FRB 20180916B \citep{Marcote20}, and 20200120E \citep{Bhardwaj21a} both of which show ${\rm RM} \sim 0$ rad m$^{-2}$. Finally, \cite{Yang26} recently showed that the scatter of the same relation can be used to constrain evolutionary models (i.e. the evolution with time of the nebula radius) for PRSs.

The YLZ relation has so far been tested only on the handful of confirmed PRS systems. A systematic, statistically meaningful test, spanning the full dynamic range of FRB RM values, from highly Faraday-rotated sources to essentially unrotated ones, and including both repeaters and one-off events, is lacking. The present work exploit the FRB--PRS catalog presented in Paper I to carry out the first systematic analysis of the $L_\nu - |{\rm RM}|$ relation across a large and well-defined sample of FRB sources.

The paper is structured as follows. In Section \ref{sec: theoretical} we outline the theoretical framework on which this work is based; in Section \ref{sec: method} we describe the properties of the sample of sources considered for the analysis and we present the constraints on the YLZ relation in Section \ref{sec:RMLnu}. Finally, we outline conclusions of this work in Section \ref{sec: conclusions}.

\section{Theoretical framework}\label{sec: theoretical}

In this Section, we outline the theoretical framework adopted throughout this work, which builds upon previous literature studies \citep{Yang20, Yang22, Bruni23, Yang26}. These studies provide a complete and rigorous model for a synchrotron emitting nebula which surrounds an FRB-emitting compact object (e.g. a magnetar). Here we only present the key ingredients of this framework that are relevant for the interpretation of our results, focusing in particular on the $L_\nu$--RM relation and its connection to the physical properties and temporal evolution of the emitting nebula.

\subsection{The $L_\nu$ - RM relation for FRB-PRS systems}

The PRS is usually modeled as a nebula having magnetic field $B$, consisting of a mixture of relativistic electrons, which emit via synchrotron radiation mechanism, and thermal electrons, the latter responsible for the bursts Faraday rotation. Following \cite{Yang26}, we consider an electron population with momentum $p$ distributed as $f(p) = n_e(p)/n_{e,0}$, where $n_e(p)$ is the number density of electrons having momentum $p$ and $n_{e,0}$ is the total electron density. Electrons have effective mass $m_e = \gamma m_{e,0}$, with
$m_{e,0} \simeq 9.11 \times 10^{-28}$ g, and $p = \sqrt{\gamma^2 - 1}$, where $\gamma$ is the Lorentz factor of the electron. An FRB passing through this population of electrons will exhibit an observable RM given by \citep{Yang20,Yang22,Bruni23,Yang26}:

\begin{equation}
    {\rm RM} = \frac{e^3}{2\pi m_{e,0}^2 \gamma_c^2 c^4} \int_0^D n_{e,0} B_\parallel {\rm d}s\ ,
\end{equation}

where $D$ is the PRS distance to the observer and $B_\parallel$ is the parallel component to the line of sight of the nebula magnetic field. Here $\gamma_c$ is the critical (or equivalent) Lorentz factor, given by \citep{Yang26}
\begin{equation}\label{eq: gamma_c}
\gamma_c \equiv \left( \int_0^{\infty} \frac{f(p)}{1 + p^2} \rm{d} p\right)^{-1/2}\ .
\end{equation}
Moreover, only electrons that radiate in the GHz band contribute to the PRS radio emission \citep{Yang20,Yang22,Yang26}. The fraction of such electrons with respect to the whole population is 

\begin{equation}\label{eq: zeta_e}
\zeta_e = \int_{p_{\rm GHz}}^\infty f(p)\ {\rm d}p\ .
\end{equation}

It can be shown that, in this framework, a spherical nebula of electrons, with radius $R$, radiating via synchrotron emission has a specific luminosity \citep{Yang20,Yang22,Bruni23,Bruni24,ZhangZhang25,Yang26}:

\begin{equation}\label{eq: RM_Lnu}
\begin{aligned}
    L_\nu &= \frac{64\pi^3}{27} \zeta_e \gamma_{\rm c}^2 m_{e,0} c^2 R^2 |{\rm RM_{\rm rest}}|\\
    &\approx 5.7 \times 10^{28}\ {\rm erg\ s^{-1}\ Hz^{-1}}\ \zeta_e \gamma_{\rm c}^2 \ \Biggl(\frac{R}{0.01\ {\rm pc}} \Biggr)^2 \Biggl(\frac{{\rm |RM_{\rm rest}|}}{10^4\ {\rm rad\ m^{-2}}} \Biggr) .
\end{aligned}
\end{equation}

In Section \ref{sec:RMLnu} we constrain the principal parameters of this relation, exploiting our extended catalog.

\subsection{Evolutionary models for PRSs}\label{sec:evol_sec}

The relativistic electrons responsible for the observed PRS emission must be accelerated through specific physical mechanisms. \citet{Yang26} discuss three main astrophysical scenarios: (i) non-relativistic shocks propagating outward (forward shocks) or inward (reverse shocks) between a supernova remnant (SNR) and the external ISM, as well as forward shocks in a pulsar wind nebula (PWN) embedded in a SNR; (ii) re-acceleration of fossil (i.e. low-energy) electrons at the termination shock in a PWN/SNR system (so-called PWN bubble); and (iii) a bow shock formed through the interaction between the wind of a neutron star and the wind of its stellar companion in a binary system. For simplicity, we restrict the considered models only to isolated progenitor scenarios, namely models (i) and (ii). These models predict different evolutionary behaviours of the radius of the emitting nebula, i.e. different $\hat{\alpha}$ values, where $R \propto t^{\hat{\alpha}}$.

 \begin{figure*}
\sidecaption
  \includegraphics[width=12cm]{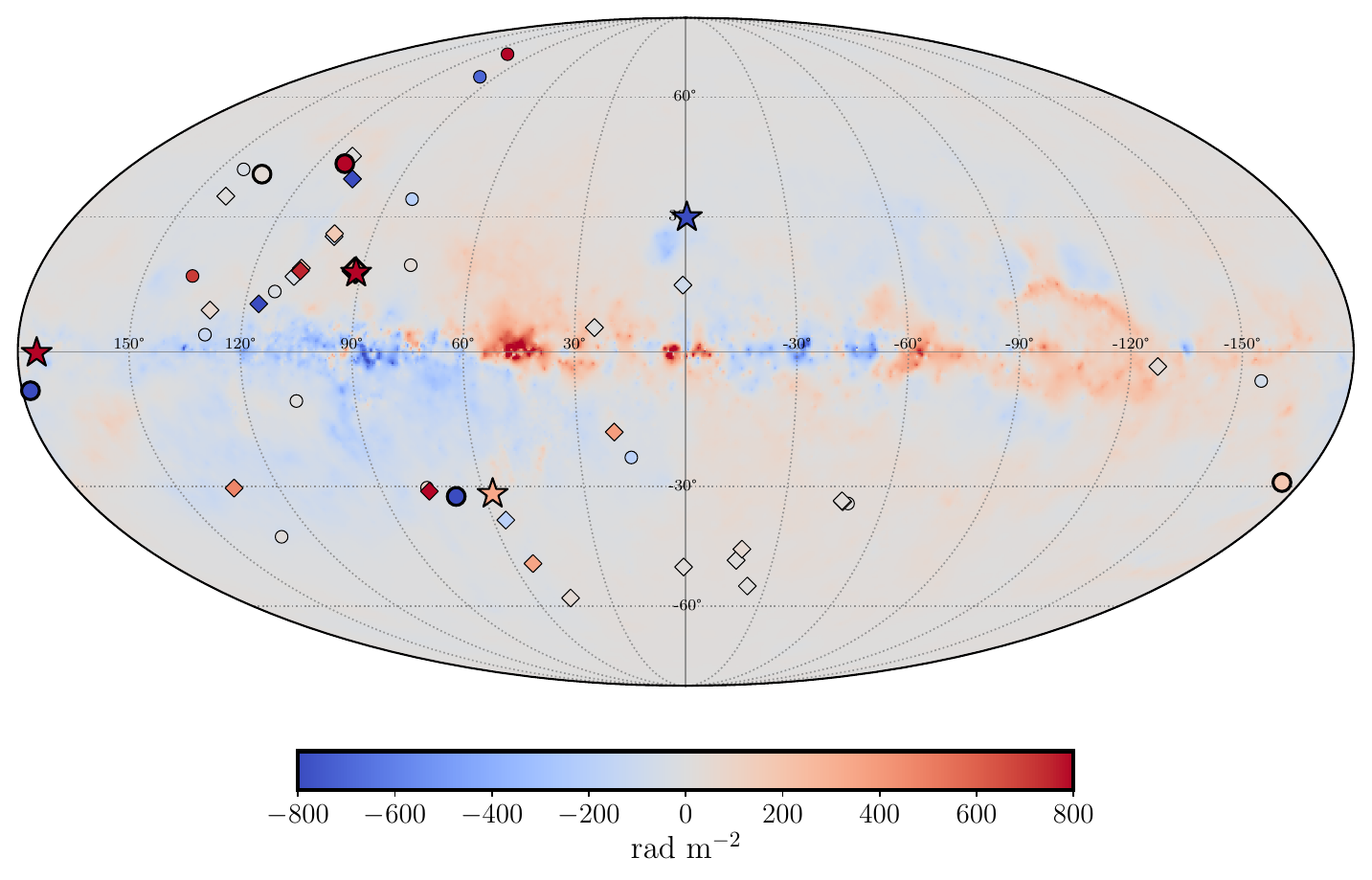}
     \caption{Mollweide projection of the all-sky Faraday RM map from \citet{Hutschenreuter21}, shown in Galactic coordinates. Overlaid are the FRB sources from our sample (see Table \ref{table:FRB_sample}). Stars indicate confirmed PRSs; circles (diamonds) with thick borders mark candidate PRSs associated with repeating (one-off) FRBs; circles (diamonds) with thin borders represent repeating (one-off) FRBs for which no persistent counterpart has been reported.}

     \label{fig: spatial_dist}
\end{figure*}

Recently, \citet{Yang26} showed that the scatter in the YLZ relation can be used to constrain this parameter. Since the composite quantity $\zeta_e \gamma_c^2$ depends only on the electron momentum distribution (see Eqs. \ref{eq: gamma_c} and \ref{eq: zeta_e}), it is expected to evolve with the physical conditions of the nebula, and hence with its size $R = R(t)$. This motivates adopting a generalized scaling of the form $L_\nu \propto R^\epsilon {\rm |RM|}$ \citep{Yang26}. It is possible to show that under some key assumptions, i.e. a steady formation rate of PRSs, such that the number of sources is proportional to their age, and a sufficiently broad size distribution for which $\ln(R_{\rm max}/R_{\rm min}) \gg \hat{\alpha}$ \citep{Yang26}, the temporal index $\hat{\alpha}$ can be expressed in terms of the scatter of the YLZ relation $\sigma_\Delta$ as  

\begin{equation}\label{eq:alpha}
    \hat{\alpha} = \ln 10\ \Bigl(\frac{\sigma_{\Delta}}{|\epsilon|}\Bigr)\ .
\end{equation}

In this framework, the observed scatter directly traces the distribution of PRS sizes within the observed population.

Different physical scenarios predict different values of the composite parameter $\hat{\alpha}|\epsilon|$, also depending on the specific evolutionary phase of the system \citep[but see][for a detailed description of the models]{Yang26}.

In non-relativistic shock waves in SNR and ISM, electrons momentum follow a Maxwell-Boltzmann distribution in the downstream, while accelerated electrons in the upstream follows a power-law distribution with slope $s$, i.e. $f(p){\rm d}p \propto p^{-s}$. It is possible to show that, being $V_{\rm sh}$ the velocity of the shock wave with respect to the upstream, and assuming a relation between the latter and the radius of the nebula, i.e. $V_{\rm sh} \propto R^{\beta_V}$, then one obtains $L_\nu \propto R^\epsilon |{\rm RM}|$, with $\epsilon = 2 + \beta_V (s-1)$. Different evolutionary phases (i.e. free-expansion or Sedov-Taylor regimes for the shock wave) are characterized by different values for $\hat{\alpha}$ and $\beta_V$, so that one can obtain the value for $\epsilon$ (and hence for $\hat{\alpha}|\epsilon|$) as a function of $s$.

 \begin{table}[ht]
\centering
\caption{List of models predictions for the composite parameter $\hat{\alpha}|\epsilon|$.}
\label{tab: models}
\begin{tabular}{lccc}
\hline
\hline
ID & Model description & $\hat{\alpha}|\epsilon| $ \\
\hline
M1 & FS in SNR/ISM, free exp. & 2.0 \\
M2 & FS in SNR/ISM, Sedov-Taylor & 0.2 \\
M3 & RS in SNR/ISM, free exp. & 3.5 \\
M4 & FS in PWN/SNR, free exp. & 2.6 \\
\hline
M5 & PWN bubble, free exp. & 1.0 \\
M6 & PWN bubble, Sedov-Taylor ($t \ll t_{\rm sd}$) & 1.0 \\
M7 & PWN bubble, Sedov-Taylor ($t \gg t_{\rm sd}$) & 0.0 \\
\hline
\end{tabular}\tablefoot{Acronyms FS, RS and free exp. denote the forward shock, reverse shock and free expansion, respectively. The predictions are shown for a fixed value of $s = 2$. Adopting $s = 3$ would result in variations of $\lesssim 10\%$ in $\hat{\alpha}|\epsilon|$ \citep[see][]{Yang26}, except for model M3, for which $\hat{\alpha}|\epsilon| = 5.0$ is predicted. The horizontal line separates models based on non-relativistic shocks and PWN bubbles (see text for details).}
\end{table}

Regarding PWN bubble models, the electron population at the TS can be separated into two components \citep[see, e.g.,][and references therein]{Yang26}: freshly injected ultrarelativistic wind electrons ($\gamma \approx 10^4 - 10^6$), and an older, cooled population ($\gamma \approx 10^2 - 10^3$) called ``relic'', which is responsible for the PRS emission. The latter is assumed to follow a power-law momentum distribution, i.e. $n(\gamma){\rm d}\gamma \propto \gamma^{-s}$. In these models, the $\hat{\alpha}$ values are determined from the evolutionary stage of the PWN/SNR bubble, as well as from the spindown of the neutron star. \citet{Yang26} show that in the free-expansion phase, $\epsilon = 1/\hat{\alpha}$ and $\hat{\alpha} = 6/5$, i.e. both independent of the index $s$. In the Sedov-Taylor regime, instead, the value of $\hat{\alpha}$ depends whether the system age of the central NS is smaller (or not) than the NS spindown timescale\footnote{this is the typical timescale over which the bulk of the initial rotational kinetic energy of the NS is released \citep[e.g.][]{Rahaman26}.} $t_{\rm sd}$. One can show that $\epsilon \sim (1-s)$.

We list in Table \ref{tab: models} all the predictions for the different models outlined in this Section. As reported in the Table, PWN bubble models (i.e. models M5-M7) generally predict lower values for $\hat{\alpha}|\epsilon|$ compared to the shock models (M1-M4), apart from model M2 which predicts $\hat{\alpha}|\epsilon| = 0.2$ for $s = 2$. 

\section{Sample description}
\label{sec: method}

The sample we use in this work is a subset of the one presented in Paper I, to which we refer the reader for the full characterization of the individual sources. Here we summarize its main properties as relevant to the analysis carried out in this paper.

The sample of Paper I comprises a total of $75$ FRB sources, of which 25 are confirmed repeaters and 50 are apparently one-off events. For each source, either a PRS flux density measurement or a $95\%$ confidence level (CL) upper limit on the spectral luminosity at $1.26$ GHz is available. The sample includes four confirmed PRSs, i.e. the ones associated with R1 \citep{Chatterjee17}, R1-twin \citep{Niu21}, 20190417A \citep{Ibik24,Moroianu26} and 20240114A \citep{Bruni24,Bhusare25}, all compact at parsec scales. Additionally, several PRS candidates are included, associated with both repeating FRBs (20201124A, 20181030A, 20201114A, 20201130A, 20230607A) and apparently one-off sources (20210317A, 20210320C, 20200906A). For these sources, co-spatial compact radio emission has been reported but their physical sizes remain unconstrained at parsec scales and/or whose FRB--PRS positional association is not yet firmly established.

Of the 75 sources, 50 have a measured RM. The latter represent the catalog of sources we consider in this work, and it is reported in Table \ref{table:FRB_sample}. The spatial distribution of these sources in Galactic coordinates is reported in Fig. \ref{fig: spatial_dist}, in which we also highlight their observed RM compared to the Galactic RM contribution map \citep{Hutschenreuter21}. 

We estimated the RM in the rest-frame of the source ${\rm RM}_{\rm rest}$ as \citep[e.g.,][]{PastorMarazuela25,ZhangZhang25}:
\begin{equation}\label{eq: RM_host}
    {\rm RM}_{\rm rest}(z) = ({\rm RM}_{\rm obs} - {\rm RM}_{\rm MW}) \times (1+z)^2,
\end{equation}
where ${\rm RM}_{\rm obs}$ is the measured RM, ${\rm RM}_{\rm MW}$ is the MW contribution to the measured RM \cite{Hutschenreuter21} and the $(1+z)^2$ term accounts for the Universe expansion. We show as well the distribution of ${\rm RM}_{\rm MW}$ in Fig. \ref{fig: spatial_dist}. The estimated RM values, as well as the Galactic RM contribution used to compute them, are listed in Table~\ref{table:FRB_sample}.

For FRBs with a time-varying RM, we adopt the maximum absolute value reported in the literature, as this better represents the extremes of the magnetized environment and is most informative for the correlation analysis (see Table~1 of Paper I) for the full list of values and references).

Spectral luminosities at $1.26$ GHz were computed assuming a spectral index $\delta = -0.27$ \citep[i.e. the spectral index measured at low frequecies for R1;][]{Marcote17} for the $k$-correction, where $F \propto \nu^\delta$. For sources without a secure host galaxy, redshifts were estimated from the observed DM via the Macquart relation \citep{Macquart20}. Finally, literature luminosities were scaled to $1.26$ GHz adopting the same spectral index (see Paper I for further details).

\begin{table}[htbp]
\centering
\caption{Peak flux density and spectral luminosity ULs for the PRS candidates in our sample.}
\label{tab: cons}
\begin{tabular}{l l c c}
\hline\hline
FRB source & ID & $F_{\rm peak}$ & $L_{1.2}$ \\
   &      & ($\mu$Jy)          & ($10^{29}\,\mathrm{erg\,s^{-1}\,Hz^{-1}}$) \\
\hline
r20201114A & S11 & $< 40$ & $<0.6$ \\
r20201130A & S12 & $<36$ & $<0.2$ \\
20210317A & S13 & $<20$ & $<1.4$ \\
r20230607A & S20 & $<30$ & $<0.6$ \\
r20181030A & - & $<10^{\rm a}$ & $<9.6 \times 10^{-5}$ \\
r20201124A & - & $<6^{\rm b}$ & $<1.4 \times 10^{-2}$ \\
\hline
\end{tabular}
\tablefoot{Flux limits adopted in the conservative scenario, in which the persistent sources are assumed not to be associated with the corresponding FRBs. Reported limits are at $95\%$ CL. \\ {\bf References}: $^{\rm a}$ \citet{Ibik24}, $^{\rm b}$ \citet{Bruni23}.}
\end{table}

As done in Paper I, to distinguish between confirmed and candidate PRSs, we will consider two different scenarios for the data analysis and interpretation: an inclusive (i.e. optimistic) one in which we treat as PRSs all the persistent radio detections, i.e. nine repeating sources (R1, R1-twin, S5, S11, S12, S20, S22, FRB 20201124A and FRB 20181030A) plus one\footnote{Also one-off sources FRBs 20210320C (S14 in Paper I) and 20200906A present persistent and co-spatial compact radio emission \citep{Mfulwane26}. However, no RM is known for these sources, hence we do not consider them in the analysis of this work.} seemingly one-off FRB (S13); and a conservative one in which only sources confirmed to be compact at milli-arcsecond angular scales (i.e. four repeaters: R1, R1-twin, S5 and S22) are considered as effective PRSs. In this latter scenario, candidate PRSs are treated as unrelated to the corresponding FRBs, and their spectral luminosities are replaced by the ULs reported in Table~\ref{tab: cons}. As in Paper I, we do not consider FRB 20220912A as a candidate PRS. This choice is based both on the non-detection of the persistent source at milliarsecond angular resolution \citep{Hewitt24} and on its radio spectrum at arcsecond scales, the latter consistent with star-formation processes \citep{Pelliciari24,Bhusare25}. 

\section{Constraints on the YLZ relation for FRB--PRS systems}\label{sec:RMLnu}

We exploit the catalog of sources of Table \ref{table:FRB_sample} to constrain the parameters entering in the YLZ relation (see Section \ref{sec: theoretical}). The latter predicts a linear relation between the bursts RM and the spectral luminosity of the associated PRS. To test the linearity of this relation we parametrize it as a power law $L_\nu \propto {\rm RM}^\beta$.
Accordingly, Eq. \eqref{eq: RM_Lnu} becomes:

\begin{equation}\label{eq: YLZ}
    L_{1.2} \approx 5.7 \times 10^{28}\ {\rm erg\ s^{-1}\ Hz^{-1}}\ \zeta_e \gamma_c^2 \ \Biggl(\frac{R}{0.01\ {\rm pc}} \Biggr)^2 \Biggl(\frac{{\rm |RM_{\rm rest}|}}{10^4\ {\rm rad\ m^{-2}}} \Biggr)^\beta .
\end{equation}

 \begin{figure}
    \centering
    \includegraphics[width=1.0\columnwidth]{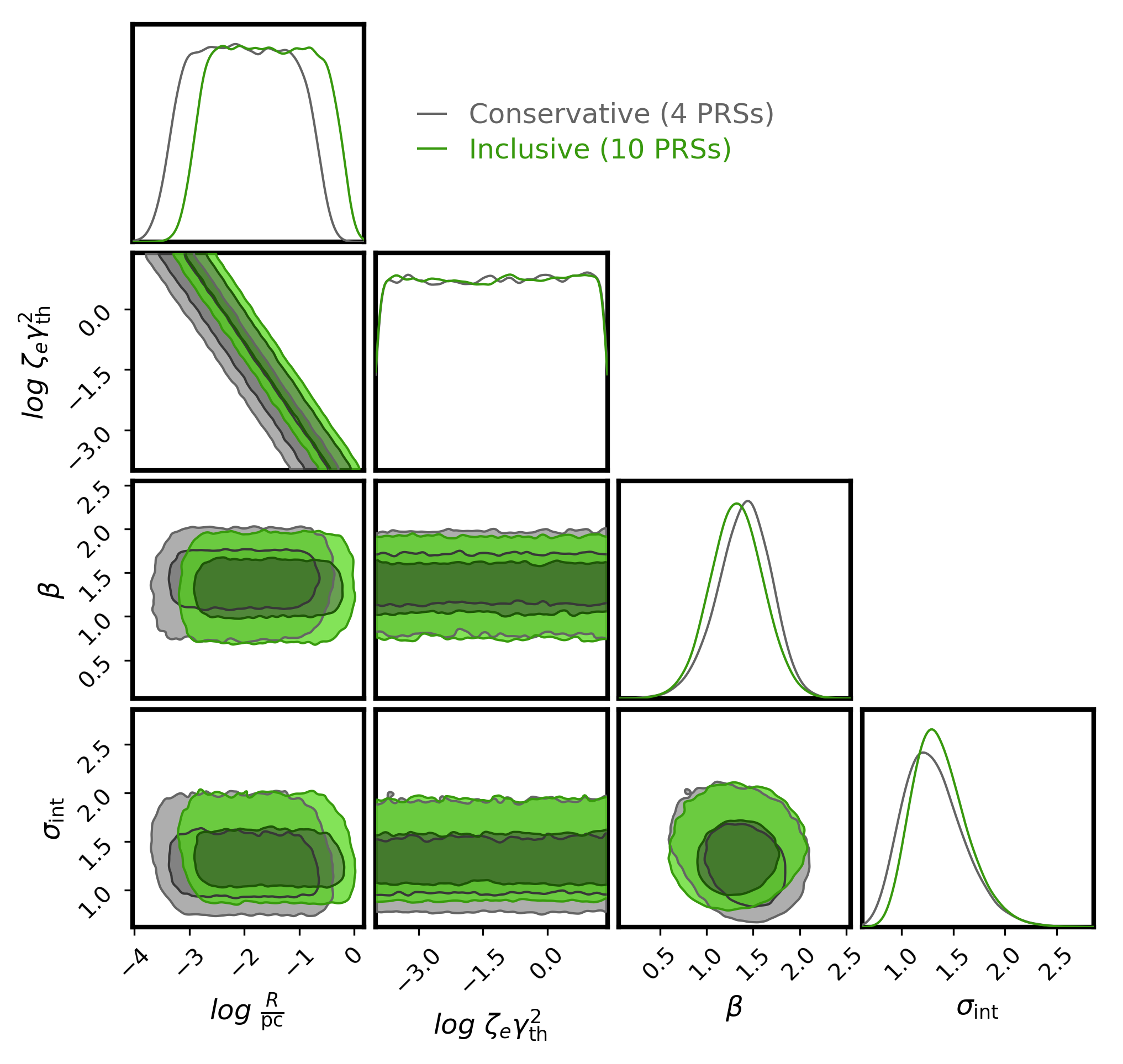}
    \caption{Posterior probability distributions of the empirical relation parameters entering in Eq. \eqref{eq: RM_Lnu}. Dark and light areas show the $68\%$ and $95\%$ confidence regions, respectively. Gray (green) contours show the posterior distributions resulting from fitting Eq. \eqref{eq: RM_Lnu} by considering a conservative (inclusive) case for the population of known PRSs (see Section \ref{sec:RMLnu}).}
    \label{fig: contours}
\end{figure}

In this analysis, we consider RM values with the correction for both the MW contribution and for the host galaxy redshift (see Eq. \ref{eq: RM_host}) already applied. We verified that the results difference in using the observed (uncorrected) RM values is less than $10\%$. In this work, we do not account for uncertainties in the RM measurements. Another potential source of systematic uncertainty is the temporal variability of the RM, either stochastic or secular, which may in turn induce variations in the observed luminosity \citep{Li26_1}. These aspects are left for future investigation.
 
 We decided to perform the fit in logarithmic space\footnote{In this work the notation $\log$ refers to base-10 logarithm.}, recasting Eq. \eqref{eq: YLZ} in terms of $\log L_{1.2}$ and $\log {\rm RM}$. We then use the following fitting formula:
\begin{equation}
\log L_{1.2} = C + \log(\zeta_e \gamma_c^2) + 2 \log R + \beta \left(\log {\rm RM_{\rm rest}} - 4 \right)\ ,
\end{equation}

with $C = \log (5.7 \times 10^{28}) - 2\log (0.01) \approx 37.966$. In this formulation, the free parameters of the model are $\log R$, $\log(\zeta_e \gamma_c^2)$, and $\beta$. We also fit a fourth parameter, i.e. $\sigma_{\rm int}$, which represents the intrinsic scatter of the best-fit. The nebula radius, $R$, is expressed in units of parsec. We then consider the dataset $\mathcal{D} \equiv (\log {\rm RM^i_{\rm rest}}, \log L_{1.2}^i)$, and follow a Bayesian approach to sample the posterior distribution of the parameters. We decided to fit for the logarithms of $R$ and $\zeta_e \gamma_c^2$, in order to properly sample the low-value regime of the parameter space. This choice allows for a more efficient and unbiased sampling of the posterior when the parameters are both intrinsically positive–definite and span a wide dynamical range. 
 
 Being $\Theta \equiv (\log R, \log  \zeta_e\gamma_c^2, \beta, \sigma_{\rm int})$ the set of free parameters, the posterior distribution $\mathcal{P}(\Theta|\mathcal{D})$ can be computed from the Bayes’s theorem, i.e. $\mathcal{P}(\Theta) \propto \mathcal{P}(\Theta)\mathcal{L}(\mathcal{D}|\Theta)$, where $\mathcal{P}(\Theta)$ is the prior distribution of $\Theta$ and $\mathcal{L}(\mathcal{D}|\Theta)$ is the likelihood function.

 \begin{figure*}
    \centering
    \includegraphics[width=0.95\textwidth]{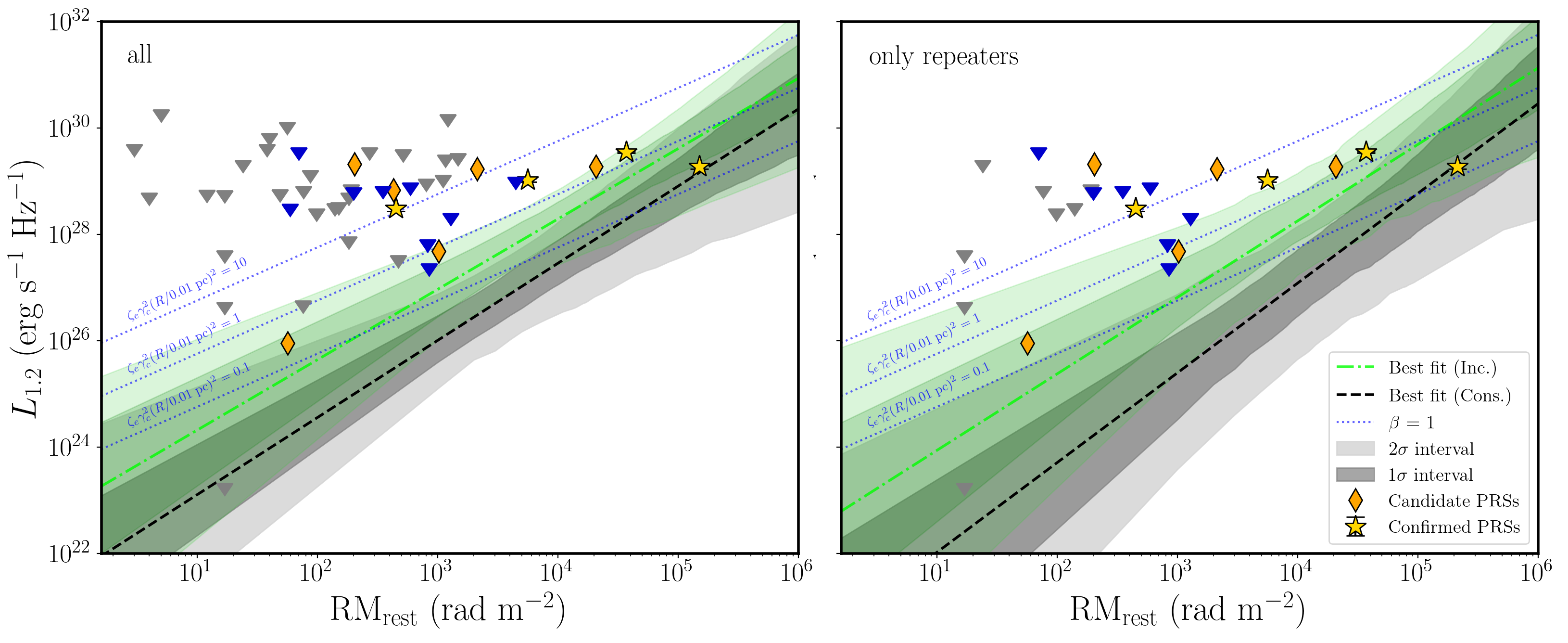}
    \caption{Relation between PRS spectral luminosity and FRB RM, the latter corrected for the MW contribution and computed in the FRB host reference frame. Yellow stars indicate confirmed PRSs (see text) while downward triangles indicate $2\sigma$ ULs to the PRS luminosity from either literature observations (gray) or our sample (blue). Orange diamonds represent candidate PRSs, which are treated as ULs in the conservative scenario. The best-fit model within the conservative (inclusive) scenario is showed as a black (green) dashed line, while associated $1\sigma$ and $2\sigma$ uncertainties on the best-fit parameters are shown as a light and dark gray (green) shaded region, respectively. The left panel shows best-fit models as obtain by including all sources in the MCMC, while only repeaters are included in the best-fit models shown in the right panel. The best fit model takes into account also ULs (see Eq. \ref{eq:likelihood}).}
    \label{fig: RMLnu}
\end{figure*}

We then make use of Markov Chain Monte Carlo (MCMC) methods for the posterior sampling, using the {\sc python} package {\sc emcee}\footnote{{\sc emcee}: \url{https://emcee.readthedocs.io/en/stable/}} which implements the Goodman--Weare posterior sampling algorithm \citep{GoodmanWeare10}. We consider flat priors on $\Theta$, with parameters free to vary in the following range: $-5.3 \leq \log R \leq 1$, $-4\leq \log \zeta_e\gamma_c^2 \leq 1.4$,  $-10 \leq \beta \leq 10$ and $0 \leq \sigma_{\rm int} \leq 4$. Regarding the prior chosen for the nebula size $R$, this converts to the range $1\ {\rm AU} \approx 4.86 \times 10^{-6}\ {\rm pc} \leq R \leq 10\ {\rm pc}$, while for the composite parameter $\zeta_e \gamma_c^2$, we consider a large range taking into account the fact that $\gamma_c \lesssim$ a few \citep{Yang22}. We considered a maximum value of $\gamma_c = 5$ so that ${\rm max} (\log \zeta_e \gamma_c^2) = \log 25 \simeq 1.4$, given that $\zeta_e$ is a fraction with $\zeta_e^{\rm max} = 1$. 
 
 Our dataset is composed by both detections, where we have a measured value for $L_{1.2}$, and non-detections, where instead we have only ULs for $L_{1.2}$. We want to fit our complete dataset, i.e. detections and non-detections, with a single likelihood function \citep{Cuciti23}:

 \begin{equation}\label{eq:likelihood}
 \begin{aligned}
     \mathcal{L} = \prod_i^{N_{\rm det}} &\frac{1}{\sqrt{2\pi(\delta_{\rm log,i}^2 + \sigma_{\rm int}^2)}}\exp\Biggl(\frac{(y_i^m - y_i)^2}{\delta_{\rm log,i}^2 + \sigma_{\rm int}^2}\Biggr) \\
     &\times \prod_i^{N_{\rm ul}} \frac{{\rm erf}\Bigl(\frac{y_i^m - Y_{\rm min}}{\sqrt{2}\sigma_{\rm int}}\Bigr) - {\rm erf} \Bigl(\frac{y_i^m - Y_{\rm max}}{\sqrt{2}\sigma_{\rm int}}\Bigr)}{2(Y_{\rm max} - Y_{\rm min})},
\end{aligned}
 \end{equation}
 
 where $y_i^m = \log L_{1.2}({\rm RM_i;\Theta})$ as predicted from Eq. \ref{eq: RM_Lnu}, while $y_i = \log L^i_{1.2}$. We fit the detection part with a standard Gaussian likelihood and the non-detection part (lower part of Eq. \eqref{eq:likelihood}) by marginalising over the true (unobserved value) of the specific luminosity for a given data-point. In particular, we consider the specific luminosity value in case of an UL to be uniformly distributed between $Y_{\rm min} = \log L_{1.2}^{\rm UL} - N_{\rm UL}$, with $N_{\rm UL} = 6$\footnote{We choose $N_{\rm UL} = 6$ rather than $N_{\rm UL} = 1$, as instead used in \cite{Cuciti23}. We tested the convergence of fitting parameters in the MCMC to a stable value for $N_{\rm UL} \geq 6$ (i.e. the fit does not depend on $N_{\rm UL}$ anymore above this value).}, and $Y_{\rm max} = \log L_{1.2}^{\rm UL}$, where $L_{1.2}^{\rm UL}$ is the luminosity value reported in the upper-limit. In Eq. \eqref{eq:likelihood}, $\delta_{\rm log}$ is the uncertainty related to $\log L_{1.2}$, i.e. $\delta_{\rm log,i} = L^i_{1.2} / (\ln 10 \ \delta_{L,i})$, and $\delta_{L,i}$ is the uncertainty associated with $L_{1.2}^i$.

We list all the MCMC results in Table \ref{tab: MCMCresults}, and we show the corresponding contour plots  for the fitting parameters in Fig. \ref{fig: contours}. As done for the rest of the analysis conducted in this work, we show the resulting contours and best-fit models in the conservative and inclusive scenarios. The obtained best-fit YLZ relations are shown in Fig. \ref{fig: RMLnu}, where we also compare the best-fit results obtained when considering both repeaters and one-off events, and those obtained considering only confirmed repeaters (right panel of Fig. \ref{fig: RMLnu}).

 \begin{table*}[ht]
\centering
\caption{MCMC results on the YLZ relation parameters.}
\label{tab: MCMCresults}
\begin{tabular}{lccccc}
\hline
\hline
Scenario & $\log (R/{1\ \rm pc})$ & $\beta$ & $\sigma_{\rm int}$ & $\hat{\alpha}|\epsilon|$ \\
\hline
all & & & & \\
Cons. & $-1.8 \pm 0.9$ & $1.43 \pm 0.26$ & $1.21^{+0.33}_{-0.26}$ & $2.8^{+0.7}_{-0.6}$ \\
Incl. & $-1.4 \pm 0.9$ & $1.35 \pm 0.28$ & $1.34^{+0.29}_{-0.23}$ & $3.1^{+0.6}_{-0.5}$ \\
Cons. ($\beta = 1$) & $-2.1 \pm 0.9$ & 1 & $1.37^{+0.32}_{-0.27}$ & $3.1^{+0.7}_{-0.6}$ \\
Incl. ($\beta = 1$) & $-1.6 \pm 0.9$ & 1 & $1.32^{+0.29}_{-0.23}$ & $3.0^{+0.7}_{-0.5}$ \\
\hline
only repeaters & & & & \\
Cons. & $-1.8 \pm 0.9$ & $1.7 \pm 0.4$ & $1.54^{+0.52}_{-0.38}$ & $3.5^{+1.1}_{-0.9}$ \\
Incl. & $-1.4 \pm 0.9$ & $1.43^{+0.40}_{-0.37}$ & $1.55^{+0.46}_{-0.33}$ & $3.5^{+1.1}_{-0.7}$ \\
Cons. ($\beta = 1$) & $-2.2 \pm 0.9$ & 1 & $1.7^{+0.5}_{-0.4}$ & $3.9^{+1.2}_{-0.9}$ \\
Incl. ($\beta = 1$) & $-1.6 \pm 0.9$ & 1 & $1.48^{+0.43}_{-0.31}$ & $3.4^{+1.0}_{-0.7}$ \\

\hline
\end{tabular}\tablefoot{The parameters uncertainties are reported at $1\sigma$ CL. The second part of the table (i.e. the one below the horizontal separating line) reports the fitting results when considering a dataset composed only by repeating sources. The last column show $\hat{\alpha}|\epsilon|$, which is a derived parameter from the scatter of the YLZ relation and is related to the temporal variation of the nebula radius $R$ (see Sect. \ref{sec: evol}). The YLZ parameter $\log \zeta_e \gamma_c^2$ is missing in this Table since it is not constrained in our analysis.}
\end{table*}

The best-fit models for the two different scenarios are characterized by slopes consistent within uncertainties, although the conservative approach systematically shifts the model normalization at low RMs toward lower luminosities because candidate PRSs are treated as ULs, resulting in a steeper, as well as lower, best fit model. The inferred relation is also broadly consistent with the range of parameters considered in the literature, i.e. $\zeta_e \gamma_c^2 (R/0.01\ {\rm pc})^2 = (0.1, 1, 10)$, which we highlight with blue dotted lines in Fig. \ref{fig: RMLnu} to facilitate a comparison with previous works \citep{Yang20, Yang22, Bruni23, Bruni24, Ibik24, Yang26}. We also repeated the analysis by fixing $\beta = 1$, as commonly assumed in the literature \citep[e.g.][]{Yang20,Yang22,Bruni23}. The corresponding best-fit models are shown in Fig. \ref{fig: RMLnu_nobeta}. In this case, the fit adjusts the normalization accordingly, yielding slightly smaller values of the nebular radius $R$, while leaving the intrinsic scatter ($\sigma_{\rm int}$) unchanged within $1\sigma$ uncertainties (see Table~\ref{tab: MCMCresults}).

Our analysis constrains the size of the PRS nebula, $R$, within its prior range, with the tightest limits obtained in the conservative scenarios. From Table \ref{tab: MCMCresults}, we derive $R = 0.016^{+0.115}_{-0.014}$ pc when $\beta$ is considered as a free parameter, and $R = 0.008^{+0.058}_{-0.007}$ pc for $\beta = 1$. This limits translate into upper bounds at $1\sigma$ level of $R \leq 0.13$ pc and $R \leq 0.07$ pc for the varying and fixed $\beta$ scenarios, respectively. Moreover, the corresponding lower bounds are $R \geq 0.002\ {\rm pc} \approx 400$ AU and $R \geq 0.001\ {\rm pc} \geq 200$ AU at $1\sigma$. We obtain shallower constraints in the inclusive scenarios, i.e. $R = 0.040^{+0.276}_{-0.035}$ pc and  $R = 0.025^{+0.174}_{-0.022}$ pc for a varying and fixed $\beta$, respectively. 

Overall, the inferred constraints are consistent at $1\sigma$ when considering either the full sample (repeaters and one-offs) or repeaters only, and are all consistent with existing limits on the size of observed PRSs, with the stringest one being $R \leq 0.7$ pc at $1\sigma$ CL for R1 \citep{Marcote17}. The composite parameter $\zeta_e \gamma_c^2$ is instead not constrained, as it is fully degenerate with $R$ (see the upper-left panel of Fig. \ref{fig: contours}).

Regarding the slope of the YLZ relation, the posterior distributions consistently favour steeper values for $\beta$, with the latter being inconsistent with the reference $\beta = 1$ at $\gtrsim 1\sigma$ for all scenarios considered. In particular, we find $\beta = 1.4 \pm 0.26$ when considering the full sample and $\beta \simeq 1.4 $-$1.7$ when restricting the dataset to repeating FRBs only (see Table~\ref{tab: MCMCresults}). In particular, for the inclusive scenario we obtain $\beta = 1.35 \pm 0.28$ for the full sample and $\beta = 1.43^{+0.40}_{-0.37}$ when considering only repeaters. The conservative scenario yields steeper slopes, with $\beta = 1.43^{+0.26}_{-0.26}$ and $\beta = 1.7 \pm 0.4$ for the two datasets, respectively. Still, the canonical linear YLZ relation is recovered for all scenarios at $2\sigma$ CL.

Finally, we report the constraints on the intrinsic scatter $\sigma_{\rm int}$ for all scenarios in Table \ref{tab: MCMCresults}, and use them to constrain models for the temporal evolution of PRSs in Sect. \ref{sec: evol}. Overall, the inferred scatter is consistent across different scenarios, with slightly (expected) higher best-fit values obtained when excluding one-off sources from the sample.

As shown in Fig. \ref{fig: RMLnu}, the PRSs associated with R1, R1-twin, and S5 lie well within the $2\sigma$ confidence region of the best-fit models in both scenarios. In contrast, the PRS associated with S22, currently the confirmed source with the lowest RM, shows a mild tension with the model at $\gtrsim 3\sigma$ confidence level. The discrepancy becomes more pronounced for PRS candidates proposed in Paper I, namely those associated with FRBs 20201114A, 20201130A, and 20210317A. However, the lack of precise localizations for these sources prevents a firm identification as genuine PRSs. On the other hand, the candidate PRSs associated with FRB 20181030A \citep{Ibik24}, FRB 20201124A \citep{Bruni23}, and FRB 20230607A (this work), despite the latter not being precisely localized, are consistent with the best-fit models within the inferred uncertainties.

Among the PRS non-detections from our uGMRT observations, the most stringent luminosity ULs are obtained for sources S2 and S3. These tight constraints are primarily driven by their relative proximity, as both sources are localized in host galaxies at redshifts $z \sim 0.06$. As evident from Fig. \ref{fig: RMLnu}, a particularly noteworthy case is FRB~20200120E, the nearest known extragalactic FRB, associated with a globular cluster in the M81 galaxy \citep{Bhardwaj21a,Kirsten22glob}. For this source, the presence of a PRS is excluded down to spectral luminosities that are six orders of magnitude lower than those observed for typical confirmed PRSs. This striking contrast may suggest the existence of multiple FRB and/or magnetar formation channels. In particular, magnetars formed through merger-induced or accretion-induced collapse of white dwarfs, both proposed formation pathways for FRB~20200120E \citep[e.g.,][]{Kirsten22glob,Kremer21,Kremer22}, may not produce persistent radio counterparts, unlike those originating from massive core-collapse events.

\subsection{Constraints on evolutionary models for PRSs}\label{sec: evol}

 \begin{figure*}
\sidecaption
  \includegraphics[width=12cm]{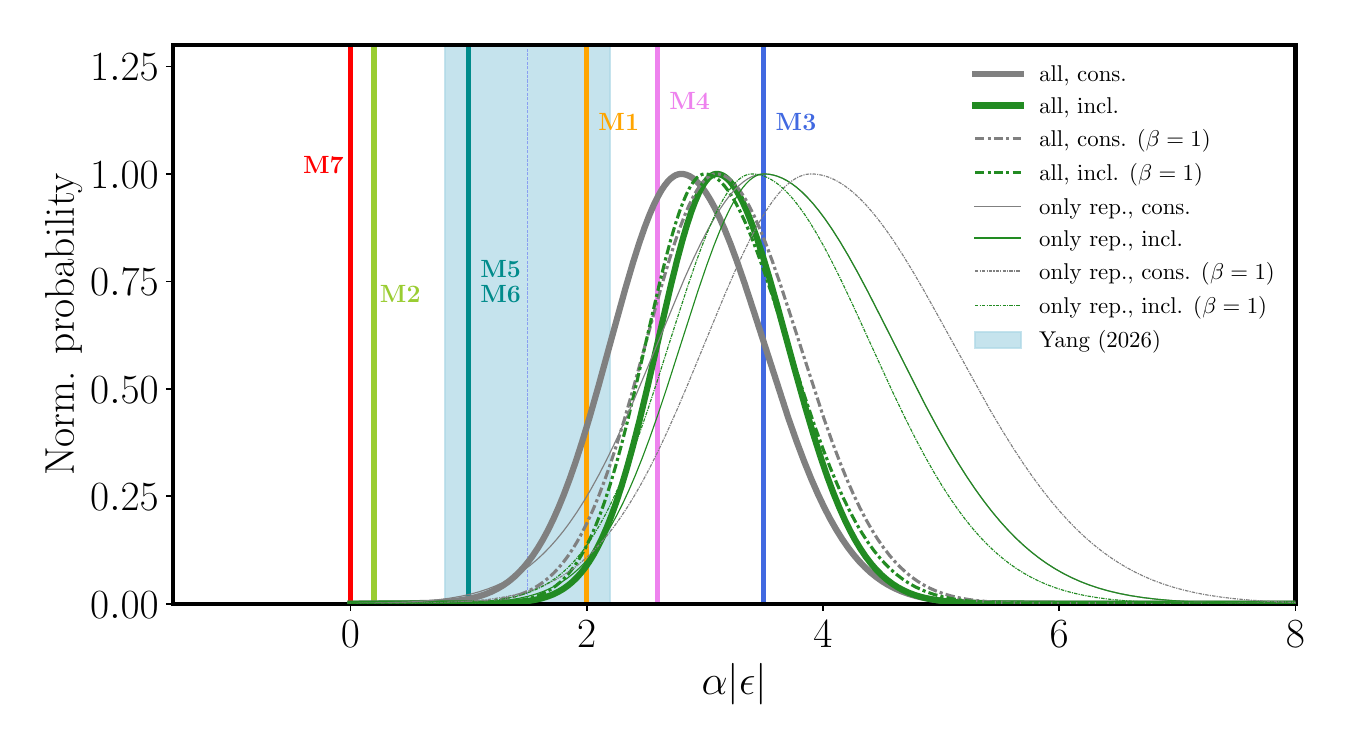}
     \caption{Probability distributions of the parameter $\hat{\alpha}|\epsilon|$, inferred from the MCMC constraints on $\sigma_{\rm int}$ (see Table~\ref{tab: MCMCresults}), for both conservative (gray) and inclusive (green) scenarios. Solid curves correspond to the case in which $\beta$ (with $L_\nu \propto |{\rm RM}|^\beta$) is treated as a free parameter in the MCMC analysis, while dash-dotted curves assume a fixed value of $\beta = 1$. Ticked curves highlight the cases where both one-off sources and repeaters are included in the fit (Sect.~\ref{sec:RMLnu}). Vertical solid lines indicate the model predictions listed in Table~\ref{tab: models}, labeled from M1 to M7. The light-blue shaded region shows the constraint on $\hat{\alpha}|\epsilon|$ reported by \cite{Yang26}.}
    \label{fig: alphaeps}
\end{figure*}

We now exploit the scatter of the YLZ relation to constrain evolutionary models for PRSs (see Sect. \ref{sec:evol_sec}). In particular we use the constraints obtained for the intrinsic scatter parameter, $\sigma_{\rm int}$, as a proxy for the composite parameter $\hat{\alpha}|\epsilon|$, where $\hat{\alpha}$ governs the temporal evolution of the nebular size, $R \propto t^{\hat{\alpha}}$, while $\epsilon$ parametrizes the generalized scaling of the $L_\nu$--RM relation with $R$, $L_\nu \propto R^\epsilon {\rm |RM|}$.

\citet{Yang26} quantifies the scatter of the YLZ relation as the standard deviation of the residuals with respect to the best-fit model obtained by fitting only confirmed PRSs, with the addition of the candidate PRS associated with FRB 20201124A. In our case, however, the presence of ULs drive the best-fit model significantly below the bulk of the detected data points. Consequently, in this case the residual-based estimate would not provide a meaningful characterization of the true dispersion of the relation. If we consider the same dataset of \citet{Yang26}, we obtain $\sigma_{\rm int} = 0.67^{+0.75}_{-0.09 }$, which can be compared to $\sigma_{\Delta} = 0.67$ \citep{Yang26}. This indicates that $\sigma_{\rm int}$ is a good proxy for the true dispersion of the YLZ relation. We then make use of Eq. \ref{eq:alpha} to convert the marginalized $\sigma_{\rm int}$ posterior distributions obtained in Sect. \ref{sec:RMLnu} into constraints for $\hat{\alpha}|\epsilon|$ in the various fitting scenarios considered (all vs. only repeaters and varying/fixed $\beta$).

The resulting distributions for $\hat{\alpha}|\epsilon|$ are shown in Fig. \ref{fig: alphaeps}, where they are compared with the theoretical expectations, namely M1--M7 models of Table \ref{tab: models}. The corresponding median values for $\hat{\alpha}|\epsilon|$, along with $1\sigma$ CL uncertainties, are   listed in Table \ref{tab: MCMCresults} for all the considered scenarios. 

We find systematically larger values of $\hat{\alpha}|\epsilon|$ with respect to $\hat{\alpha}|\epsilon| = 1.5 \pm 0.7$ reported by \citet{Yang26}, favouring evolutionary scenarios in which the emitting region undergoes rapid expansion (i.e. larger $\hat{\alpha}$ values) and/or efficient energy injection into the relativistic electron population. Models predicting low values of $\hat{\alpha}|\epsilon|$ are generally disfavoured by our analysis. In particular, the Sedov–Taylor forward shock scenario in SNR/ISM environments (M2, $\hat{\alpha}|\epsilon| \simeq 0.2$), as well as PWN bubble models in both free-expansion and Sedov–Taylor phases (M5–M7, $\hat{\alpha}|\epsilon| \lesssim 1$), lie systematically below the bulk of the inferred distributions. While marginal consistency at the $\sim 2\sigma$ level cannot be excluded, these scenarios do not naturally reproduce the large scatter observed in the data.

Conversely, forward shocks in SNR/ISM during the free-expansion phase (M1, $\hat{\alpha}|\epsilon| \simeq 2.0$) and forward shocks in PWN/SNR systems (M4, $\hat{\alpha}|\epsilon| \simeq 2.6$) are broadly consistent with the peak of the obtained distributions. Even higher values, such as those predicted for reverse shocks in SNR/ISM environments (M3, $\hat{\alpha}|\epsilon| \gtrsim 3.5$), are also compatible with the upper tails of the posteriors. However, adopting a steeper electron spectrum ($s = 3$) for the M3 scenario leads to a prediction of $\hat{\alpha}|\epsilon| \simeq 5$ \citep{Yang26}, which is in tension with our main constraints at the $\sim 2\sigma$ level. This value is only marginally compatible with the high $\hat{\alpha}|\epsilon|$ tail obtained in the specific case where the fit is restricted to repeaters and the slope is fixed to $\beta = 1$.

However, an important caveat must be emphasized. In this work, detections and ULs are treated jointly within a single population of FRB-PRS systems, while in reality the observed sample may be heterogeneous. In particular, FRBs hosting confirmed PRSs may represent a distinct subset with different physical properties and evolutionary stages. As a consequence, analyses restricted to larger samples of securely identified PRSs could yield systematically different (and potentially lower) values of $\hat{\alpha}|\epsilon|$, more in line with previous estimates based on confirmed sources alone. For this reason, the constraints presented here should be regarded as indicative. While they provide evidence in favour of high-$\hat{\alpha}|\epsilon|$ scenarios, they also highlight the need for larger and more homogeneous samples of confirmed PRSs.

An additional caveat is that the scatter of the YLZ relation could also be driven by differences in the physical properties of the underlying systems, rather than solely by their different evolutionary stages. Variations in the magnetar birth properties, energy budget, magnetic field, and surrounding environment may contribute to the observed dispersion in the $L_{\nu}$--RM plane. The mapping between $\sigma_{\rm int}$ and $\hat{\alpha}|\epsilon|$ should therefore be regarded as model-dependent, as it assumes that the observed scatter is predominantly evolutionary. These effects will be investigated in greater detail in future work.

\section{Summary and conclusions}\label{sec: conclusions}

In a companion work (Paper I) we have presented an extended sample of FRB sources, comprising new $1.26$ GHz uGMRT observations of a sample of $24$ FRBs ($13$ are confirmed repeaters and $11$ seemingly one-off sources) and literature data, obtaining a catalog of 75 FRBs for which a PRS search has been conducted. We already used it to study the occurrence of PRSs in the observed population of FRBs. In this work we exploited a subset of this catalog to deeply investigate the expected correlation between the FRBs RM and the radio spectral luminosity of the associated PRSs \citep[YLZ relation; e.g.,][]{Yang20,Bruni23}. In particular, we selected $50$ individual FRB sources with known RM. We also provided constraints for evolutionary scenarios of FRB-PRS systems, following the framework delineated in \cite{Yang26}. Being the majority of the PRS detections reported classified as candidates, we conducted our analysis under two scenarios: a conservative one, in which only confirmed PRSs are considered (4 repeaters and 0 one-offs), and an inclusive one, in which candidate PRSs are also included (9 repeaters and 3 one-offs).

We constrained the principal parameters of the YLZ relation using a Bayesian MCMC framework that accounts for both detections and spectral luminosity ULs. We considered a fitting relation of the form $L_\nu \propto \zeta_e \gamma_c^2 R^2 |{\rm RM}|^\beta$, obtaining constraints on the size of the PRS nebula, $R$, and the RM scaling index $\beta$. Our analysis permitted to obtain $R = 0.016^{+0.115}_{-0.014}$ pc at $1\sigma$ CL when considering the whole dataset and a varying $\beta$ parameter. In a more canonical scenario in which $\beta$ is kept fixed at $\beta = 1$, the size constraint approximately halves to $R = 0.008^{+0.058}_{-0.007}$ pc. Regarding the dependence of the YLZ relation with the RM, our current dataset prefers mildly steeper $\beta$ values than the canonical linear scaling ($\beta = 1$). In particular, the steepest relation ($\beta = 1.7 \pm 0.4$ at 1$\sigma$ CL) is obtained when we restrict the analysis only to repeaters, in the conservative scenario. Nevertheless, the constraints for $\beta$ are consistent within $2\sigma$ uncertainties with the canonical $\beta = 1$ scenario. 

We also considered the intrinsic scatter of the YLZ relation as a free parameter, given the large set of spectral luminosities upper limits and small number of PRS detections in our catalog. We found that the YLZ relation shows a substantial intrinsic scatter, significantly larger than previous estimates based on confirmed PRSs alone and we used the latter to investigate evolutionary models for PRSs following the prescriptions of \citet{Yang26}. The inferred scatter translates into relatively large values of $\hat{\alpha}|\epsilon|$, i.e. a composite parameter encapsulating both the temporal dependence of $R$ and the dependence of the $L_\nu$--RM relation on $R$, favouring models characterized by efficient particle acceleration and/or rapid nebular expansion. In particular, forward-shock scenarios in SNR/ISM or PWN/SNR systems are broadly consistent with the data, while models predicting low $\hat{\alpha}|\epsilon|$ (e.g. PWN bubble models or Sedov-Taylor shocks) are disfavoured. 

Future extensions of this analysis will benefit from the discovery and confirmation of additional PRSs, particularly at low RM, where the current sample provides shallow constraints on $\beta$. An interesting example is provided by the candidate PRS associated with \citep{Ibik24}, recently shown to be compact on sub-parsec scales (Paper II), although its association with the FRB remains uncertain due to the still insufficient localization precision. Such systems may reveal PRSs with physical properties that differ substantially from those of the currently confirmed population. Systematic searches in large FRB samples, such as the recent CHIME catalogue \citep{CHIMECat2}, will therefore be key to understand the properties of FRB-PRS systems.

\begin{acknowledgements}
We thank the staff of the GMRT that made these observations possible. GMRT is run by the National Centre for Radio Astrophysics of the Tata Institute of Fundamental Research. The research activities described in this paper were carried out with contribution of the NextGenerationEU funds within the National Recovery and Resilience Plan (PNRR), Mission 4 - Education and Research, Component 2 - From Research to Business (M4C2), Investment Line 3.1 - Strengthening and creation of Research Infrastructures, Project IR0000026 – Next Generation Croce del Nord. OMS's research is supported by the South African Research Chairs Initiative of the Department of Science, Technology and Innovation and the National Research Foundation (grant No. 81737).
\end{acknowledgements}

\bibliographystyle{aa}
\bibliography{biblio2}

\begin{appendix}
\nolinenumbers

\section{The catalogue}

\onecolumn
\begin{center}
\begin{longtable}{clcccc}
\caption{Properties of the FRB sample presented in this work.} \label{table:FRB_sample} \\

\hline\hline
ID & Source name & ${\rm RM}_{\rm rest}$ & ${\rm RM}_{\rm MW}$ & $L_{1.2}$ & Refs. \\
& & (rad m$^{-2}$) & (rad m$^{-2}$) & ($10^{29}$ erg s$^{-1}$ Hz$^{-1}$) & \\
\hline
\endfirsthead

\multicolumn{6}{c}%
{\tablename\ \thetable\ -- Continued} \\
\hline\hline
ID & Source name & ${\rm RM}_{\rm rest}$ & ${\rm RM}_{\rm MW}$ & $L_{1.2}$ & Refs. \\
& & (rad m$^{-2}$) & (rad m$^{-2}$) & ($10^{29}$ erg s$^{-1}$ Hz$^{-1}$) & \\
\hline
\endhead

\hline \multicolumn{6}{r}{} \\
\endfoot

\hline
\endlastfoot

S1 & r20180301A & $-200(40)$ & $+44(19)$ & $<0.75$ & ($1,2$, this work) \\
S2 & r20180814A & $+847(20)$ & $-42(16)$ & $<2.8 \times 10^{-2}$ & ($3,4$, this work) \\
S3 & $^{\rm a}$r20190303A & $-821(5)$ & $+22(4)$ & $<0.08$ & ($3,4$, this work) \\
S5 & r20190417A$^{**}$ & $+6441(24)$ & $+30(19)$ & $1.05(8)$ & ($3-6$, this work) \\
S7 & r20190804E & $-351^{+55}_{-15}$ & $+12(6)$ & $<0.8$ & ($7,8$, this work) \\
S8 & r20191106C & $+1280(3)$ & $-2(3)$ & $<0.25$ & ($7-9$, this work) \\
S10 & 20200216A & $+4450^{+854}_{-125}$ & $-36(10)$ & $<1.2$ & ($10$, this work) \\
S11 & r20201114A$^*$ & $+2130^{+70}_{-470}$ & $+8(12)$ & $1.6(7)$ & ($7,9$, this work) \\
S12 & r20201130A$^*$ & $+203^{+30}_{-40}$ & $+32(21)$ & $2.1^{+0.9}_{-1.2}$ & ($7,9$, this work )\\
S13 & 20210317A$^*$ & $+430^{+84}_{-45}$ & $+26(20)$ & $0.68(8)$ & ($10$, this work) \\
S16 & 20210117A & $+59(16)$ & $+3(9)$ & $<0.37$ & ($1,11$, this work) \\
S20 & r20230607A$^*$ & $-20700^{+3180}_{-640}$ & $-15(8)$ & $1.9(8)$ & ($12$, this work) \\
S21 & r20230814A & $-70(50)$ & $+7(18)$ & $<3.9$ & ($13,14$, this work) \\
S22 & r20240114A$^{**}$ & $+450(13)$ & $-15(10)$ & $0.31(4)$ & ($15,16$, this work) \\
S24 & r20240619D & $+592^{+25}_{-90}$ & $-28(9)$ & $<0.94$ & ($17,18$, this work) \\

\hline
$-$ & 20110523A & $-514(130)$ & $+15(10)$ & $<3.9$ & ($19,20$) \\
R1 & $^{\rm a}$r20121102A$^{**}$ & $+1.5 \times 10^5$ & $-18(37)$ & $1.9(2)$ & ($21-23$)\\
$-$ & 20150215A & $-87(90)$ & $+32(28)$ & $<1.6$ & ($24$) \\
$-$ & 20150418A & $+40(150)$ & $+52(26)$ & $<7.9$ & ($25$) \\
$-$ & 20150807A & $-3(8)$ & $+14(6)$ & $<4.9$ & ($26$) \\
$-$ & 20180309A & $<200$ & $+11(6)$ & $<1.6$ & $(20,27)$ \\
R3 & $^{\rm a}$r20180916B & $-17(42)$ & $-98(39)$ & $<3.5 \times 10^{-3}$ & ($28,29$) \\
$-$ & 20180924B & $-4(9)$ & $+16(5)$ & $<0.6$ & ($30$) \\
$-$ & r20181030A$^*$ & $+56(7)$ & $-20(7)$ & $9(1) \times 10^{-4}$ & (Paper II, $31$) \\
$-$ & 20181112A & $-12(13)$ & $+16(6)$ & $<0.68$ & ($32-34$)\\
$-$ & 20190102C & $+139(16)$ & $+27(8)$ & $<0.38$ & ($32,34,35$) \\
$-$ & r20190117A & $+190^{+32}_{-16}$ & $-29(9)$ & $<0.85$ & ($3,5$) \\
$-$ & $^{\rm a}$r20190208A & $+77(30)$ & $+4(12)$ & $<0.8$ & ($3-5$) \\
R1-twin & r20190520B$^{**}$ & $-3.6 \times 10^4$ & $-13(9)$ & $3.4(1)$ & ($36,37$) \\
$-$ & 20190608B & $+472(17)$ & $-24(13)$ & $<4 \times 10^{-2}$ & ($34,38,39$) \\
$-$ & 20190611B & $-17(22)$ & $+29(11)$ & $<0.67$ & ($6,35,38$) \\
$-$ & r20190711A & $-24(16)$ & $+19(6)$ & $<2.5$ & ($38,40$) \\
$-$ & 20191001A & $+49(7)$ & $+24(4)$ & $<0.7$ & ($41$) \\
$-$ & 20191108A & $+1210(320)$ & $-63(13)$ & $<18$ & ($42$) \\
$-$ & $^{\rm a}$r20200120E & $-17(15)$ & $-17(5)$ & $<2.1 \times 10^{-6}$ & ($43-45$) \\
$-$ & $^{\rm a}$r20201124A$^*$ & $-1020(30)$ & $-44(24)$ & $4.4 \times 10^{-3}$ & ($46-48$) \\
$-$ & 20220207C & $+182(18)$ & $-5(17)$ & $<9 \times 10^{-2}$ & ($49, 50$) \\
$-$ & 20220307B & $-1470(50)$ & $-4(29)$ & $<3.3$ & ($49,50$) \\
$-$ & 20220310F & $+56(16)$ & $-14(7)$ & $<13$ & ($49,50$) \\
$-$ & 20220319D & $+76(23)$ & $-14(18)$ & $<1.3 \times 10^{-3}$ & ($49,50$) \\
$-$ & 20220418A & $-5(40)$ & $+8(14)$ & $<22$ & ($49,50$) \\
$-$ & 20220506D & $-38(26)$ & $-10(15)$ & $<4.9$ & ($49,50$) \\
$-$ & 20220509G & $-150(18)$ & $+16(15)$ & $<0.4$ & ($49,50$) \\
$-$ & r20220529A & $+98(142)$ & $-53(7)$ & $<0.3$ & ($51$) \\
$-$ & 20220717A & $+801(33)$ & $-46(18)$ & $<1.1$ &  ($35,52$) \\
$-$ & 20220825A & $+1160(28)$ & $-3(16)$ & $<3.1$ & ($49,50$) \\
$-$ & 20220905A & $-182(60)$ & $0(25)$ & $<0.6$ &  ($35,52$) \\
$-$ & r20220912A & $+17(14)$ & $-15(11)$ & $<4.9 \times 10^{-2}$ & ($53-55$) \\
$-$ & 20220920A & $-1110(21)$ & $+1(13)$ & $<1.3$ & ($49,50$) \\
$-$ & 20221012A & $+270(36)$ & $+3(12)$ & $<4.3$ & ($49,50$) \\

\end{longtable}
\tablefoot{Columns are, from left to right, the FRB ID as adopted in the text, its Transient Name Server (TNS) identifier, with confirmed repeaters identified with an `r'' at the beginning of the name, the redshift of the FRB host galaxy, the reference for its localization and redshift, its rest-frame RM with corresponding reference, the flux density at the FRB localization, and the corresponding spectral luminosity. Flux and luminosity limits are reported at $95\%$ C.L. The table is divided in sources for which we report new observations in this work and observations from the literature. FRB sources for which a co-spatial persistent emission has been found are denoted with a $^*$ supscript, while already confirmed PRSs are denoted with $^{**}$. \\
\textbf{References}. (1) \cite{Bhandari22}, (2) \cite{Kumar23}, (3) \cite{Michilli23}, (4) \cite{McKinven23}, (5) \cite{Ibik24}, (6) \cite{Moroianu26}, (7) \cite{chime23}, (8) \cite{Ibik24_b}, (9) \cite{Ng25}, (10) \cite{PastorMarazuela25}, (11) \cite{Woodland24}, (12) \cite{Zhou25}, (13) \cite{Ravi23_ATEL}, (14) \cite{Connor24}, (15) \cite{Snelders24}, (16) \cite{Tian24}, (17) \cite{Tian25}, (18) \cite{Shaji26}, (19) \cite{Masui15}, (20) \cite{Bruni23}, (21) \cite{Chatterjee17}, (22) \cite{Hilmarsson21}, (23) \cite{Tendulkar21}, (24) \cite{Petroff17}, (25) \cite{Keane16}, (26) \cite{Ravi16}, (27) \cite{Oslowski19}, (28) \cite{Marcote20},  (29) \cite{CHIME18}, (30) \cite{Bannister19}, (31) \cite{Bhardwaj21}, (32) \cite{Heintz20}, (33) \cite{Prochaska19}, (34) \cite{Bhandari20b}, (35) \cite{Mfulwane26}, (36) \cite{Niu21}, (37) \cite{AnnaThomas23}, (38) \cite{Day19}, (39) \cite{Macquart20}, (40) \cite{Chibueze21}, (41) \cite{Bhandari20a}, (42) \cite{Connor20}, (43) \cite{Bhardwaj21a}, (44) \cite{Kirsten22glob}, (45) \cite{Koss22}, (46) \cite{Kilpatrick21}, (47) \cite{Nimmo21}, (48) \cite{Xu2022}, (49) \cite{Law23}, (50) \cite{Sherman24}, (51) \cite{Phandi26}, (52) \cite{Rajwade24}, (53) \cite{ravi23}, (54) \cite{Zhang23}, (55) \cite{Hewitt24}. \\

$^{\rm a}$The RM of these FRBs is reported to vary with time \citep[see e.g.,][and references therein]{McKinven23}. We take the maximum absolute values reported in the associated reference. An exception is FRB 20220529 for which the source present a stable RM (although with a large scatter) with a sudden RM increase to $\sim 2000$ rad m$^{-2}$ \citep{Phandi26,Li26}. For this particular case we considered the RM in the stable regime.}
\end{center}

\twocolumn

\section{YLZ relation for a fixed RM exponent}
\begin{figure*}
    \centering
    \includegraphics[width=1.0\textwidth]{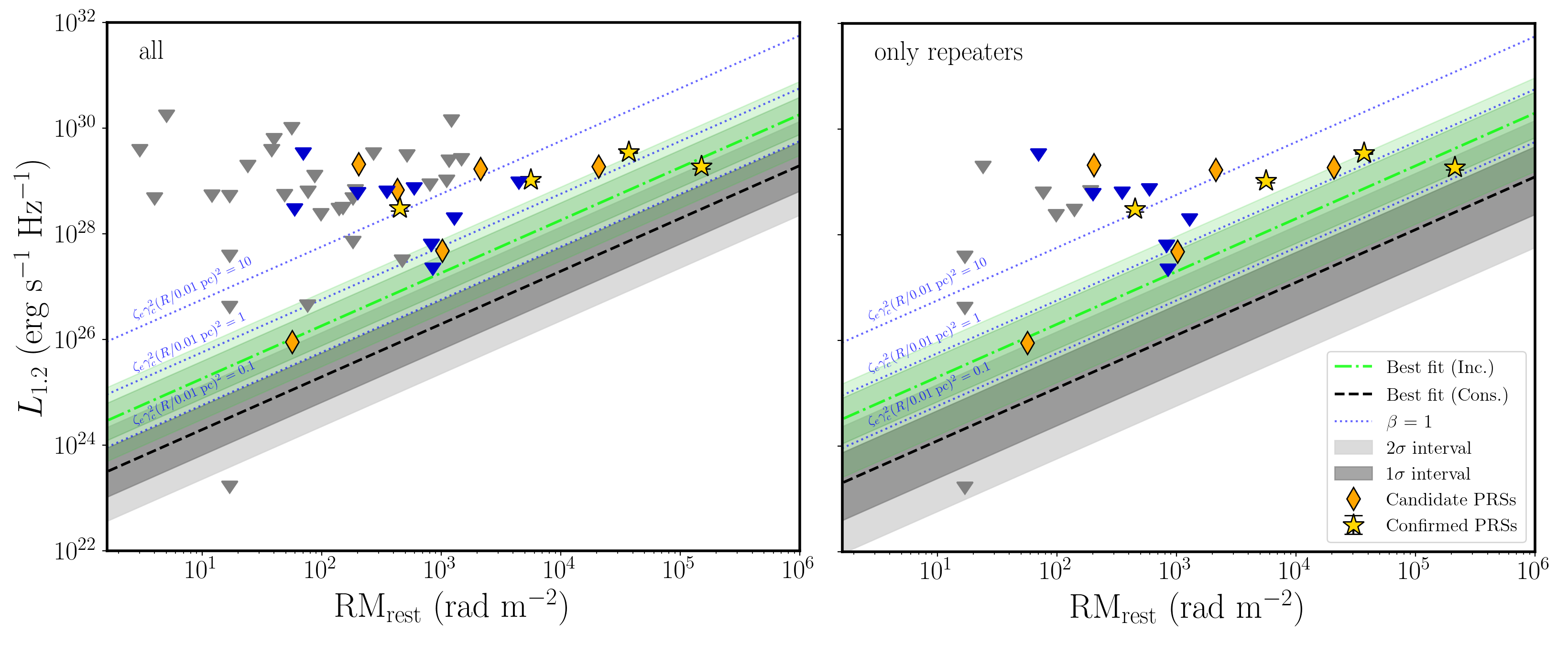}
    \caption{As Figure \ref{fig: RMLnu}, but considering $\beta = 1$.}
    \label{fig: RMLnu_nobeta}
\end{figure*}

Fig. \ref{fig: RMLnu_nobeta} shows the RM-$L_\nu$ relation obtained from the MCMC fit we described in Section \ref{sec:RMLnu} when considering the exponent of the RM term (see Eq. \ref{eq: RM_Lnu}) fixed to $\beta = 1$. 

\end{appendix}

\end{document}